%% file: relativisticMirrorResub.tex
\documentclass[reprint,longbibliography,aip]{revtex4-1}

\usepackage{graphicx} 
\usepackage{amsmath}
\usepackage{ amssymb }

\usepackage{orcidlink}
\usepackage{hyperref}
\usepackage{color}
\hypersetup{colorlinks, citecolor=cyan, linkcolor=cyan, urlcolor=cyan}

\usepackage{float} 
\usepackage{wrapfig} 

\usepackage[makeroom]{cancel} 
\usepackage{comment}

\usepackage{lipsum} 

\usepackage{outlines}

\graphicspath{{./figs/}} 

\newcommand{\beq}{\begin{equation}}
\newcommand{\eeq}{\end{equation}}

\newcommand{\bvec}{\begin{pmatrix}}
\newcommand{\evec}{\end{pmatrix}}
\newcommand{\lp}{\left(}
\newcommand{\rp}{\right)}
\newcommand{\pa}[2]{\frac{\partial #1}{\partial #2}}

\newcommand{\ve}[1]{\mathbf{#1}}

\AtBeginDocument{%
\newwrite\bibnotes
\def\bibnotesext{Notes.bib}
\immediate\openout\bibnotes=\jobname\bibnotesext
\immediate\write\bibnotes{@CONTROL{REVTEX41Control}}
\immediate\write\bibnotes{@CONTROL{%
		aip41Control,author="08",editor="1",pages="1",title="0",year="1"}}
\if@filesw
\immediate\write\@auxout{\string\citation{aip41Control}}%
\fi
}%

\begin{document}



\title{Confinement Time and Ambipolar Potential in a Relativistic Mirror-Confined Plasma}
\author{Ian E. Ochs\orcidlink{0000-0002-6002-9169}} 
\email{iochs@princeton.edu}
\author{Vadim R. Munirov\orcidlink{0000-0001-6711-1272}}
\author{Nathaniel J. Fisch\orcidlink{0000-0002-0301-7380}}

\affiliation{Department of Astrophysical Sciences, Princeton University, Princeton, New Jersey 08540, USA}

\date{\today}

\begin{abstract}

Advanced aneutronic fusion fuels such as proton-Boron$^{11}$ tend to require much higher temperatures than conventional fuels like deuterium-tritium.
For electrons, the bulk plasma temperature can approach a substantial fraction of the rest mass.
In a mirror confinement system, where the electrons are confined by an ambipolar potential of at least five electron temperatures, the tail electrons which can escape the potential are fully relativistic, which must be taken into account in calculating their confinement.
In this paper, simple estimates are employed to extend the scaling of the confinement time into the relativistic regime.
By asymptotically matching this scaling to known solutions in the non-relativistic limit, accurate forms for the confinement time (and thus the the ambipolar potential) are obtained.
These forms are verified using finite-element-based Fokker-Planck simulations over a wide range of parameters.
Comparing relativistic and nonrelativistic mirror-confined plasmas with the same ratio of confining potential $|e\phi|$ to electron temperature $T_e$ and the same mirror ratio $R$, the net result is a decrease in the confinement time due to relativistic effects by a factor of $S \equiv (1+15T_e/8m_ec^2)/(1+2|e\phi|/m_ec^2)$.


\end{abstract}

\maketitle

\section{Introduction}

Recently, there has been a revival in interest in the mirror approach\cite{Post1987,Bagryansky2019EncouragingResults} to fusion energy, which was largely abandoned with the demise of the Lawrence Livermore mirror fusion program in the 1980s.
While this renewed interest has been driven by several factors, the most important perhaps is the introduction of sheared-flow-stabilization, allowing current mirror concepts to maintain axissymmetry while avoiding the magnetohydrodynamic flute instabilities that plagued early axisymmetric mirror devices. \cite{Bekhtenev1977CertainFeatures,Huang2001}
This shear stabilization has been demonstrated to dramatically improve plasma confinment on GAMMA-10, \cite{Cho2005} the Maryland Centrifugal Mirror eXperiment (MCMX), \cite{Ellis2005} and the Gas Dynamic Trap (GDT). \cite{Beklemishev2010VortexConfinement,Ivanov2013GasdynamicTrap}
Furthermore, the associated rapid rotation of the plasma helps to improve ion confinement through centrifugal forces,\cite{Bekhtenev1980, Ellis2001,Teodorescu2010} which also reduces the phase space hole that drives kinetic loss-cone instabilities.
Combined with more efficient methods of sustaining electron temperatures in tandem mirror end-plugs,\cite{Fowler2017NewSimpler} these experiments have paved the way for the next generation of axisymmetric mirror experiments, including multiple mirror (MM) traps, \cite{Burdakov2016MultiplemirrorTrap,Beery2018PlasmaConfinement,Miller2021RateEquations} the Centrifugal Mirror Fusion eXperiment (CMFX), \cite{White2018} and the Wisconsin High-field Axisymmetric Mirror (WHAM). \cite{Egedal2022FusionBeam}

At the same time, as breakeven deuterium-tritium (DT) fusion becomes a reality, it makes sense to look forward towards advanced aneutronic fusion fuels such as proton-Boron$^{11}$ (p-B11), which---while technologically much more challenging---embody fusion's promise as a clean, abundant, nonradiactive power source much more fully than the fast-neutron producing, tritium-reliant DT reaction.
The p-B11 reaction, in particular, has been revisited by several groups,\cite{Volosov2006ACT,Volosov2011Problems,Hay2015Ignition,Putvinski2019,kolmes2022waveSupported,Ochs2022ImprovingFeasibility,Magee2019DirectObservation,Eliezer2020NovelFusion,Ruhl2022LaserPB11,Istokskaia2023MultiMeVAlpha,Magee2023FirstMeasurements} partly thanks to more optimistic fusion cross sections \cite{Sikora2016CrossSection} that improved its outlook compared to earlier pessimistic predictions. \cite{Rider1995,Nevins2000CrossSection}
While this reaction produces some neutrons due to undesirable side reactions, these neutrons have lower energy and much lower flux than those from a DT fusion reaction.

In contrast to DT, p-B11 fusion takes place at much higher temperatures; typically 300 keV for ions, and 160 keV for electrons.
Thus, relativistic effects which were ignored for DT become important, and some of the fundamental results of the mirror physics literature have to be revisited.
Here, we revisit one of the most important of these results: the relationship between the confining potential and the confinement time.

In a magnetic mirror, both ions and electrons are confined by the mirror force that results from the conservation of the magnetic moment, and are lost when they scatter into the loss cone via collisions.
However, due to their relatively low mass, the electrons collide faster, and (if the mirror plasma is rotating) are unaffected by centrifugal forces.
Thus, the electrons will leave the mirror more quickly than the ions, causing the mirror plasma to charge positive.
As was first pointed out by Kaufman,\cite{Kaufman1956} this charging will continue until the total loss rate of charge from the system, due to both ion and electron losses, goes to zero, i.e., when:
\begin{align}
    \sum_s Z_s n_s \tau_{Cs}^{-1} = 0,   
\end{align}
where for speceis $s$, $Z_s$ is the charge state, $n_s$ is the density, and $\tau_{Cs}$ is the confinement time.
The resulting potential is thus known as the ambipolar potential, and is critical in understanding the overall mirror equilibrium.

To be able to calculate the ambipolar potential and the overall performance of the mirror as a confinement device, it is necessary to be able to calculate the confinement time as a function of the mirror ratio and confining potential.
Due to its importance, continued refinements were made to Kaufman's original estimate, leading to a series of increasingly accurate approximations to the solution of the collisional momentum-space diffusion equation by Pastukhov,\cite{Pastukhov1974} Cohen et. al,\cite{Cohen1978,Cohen1980} and Najmabadi et. al,\cite{Najmabadi1984} in addition to related approaches by other authors. \cite{Budker1958,Volosov1981,Catto1985}
Generally, these approaches relied on solving the equation for a simple source and sink, and then matching these solutions as closely as possible to the shape of the loss cone.

In this paper, we extend the approximate solutions for the confinement time to relativistic plasmas.
Rather than solving the collisional diffusion equation directly, we derive the scaling of the loss rate as a combination of the perpendicular momentum diffusion timescale and the fraction of the particles with sufficient energy to escape the potential. 
We then match this scaling to the nonrelativistic solution from Ref.~\onlinecite{Najmabadi1984}, providing an accurate estimate for the loss rate, as we confirm with finite-element simulations of the full diffusion equation.

The outline of the paper is as follows.
We start in Sec.~\ref{sec:diffusionEquation} by presenting the relativistic momentum-space diffusion equation and transforming it to units recognizable from the earlier mirror literature.
In Sec.~\ref{sec:lossCone}, we do the same for the equation that determines the loss cone boundary.
Then, in Sec.~\ref{sec:analyticalEstimates}, we use these equations and the existing approximate solutions from the literature to derive analytical estimates for the confinement time.
We thus find that, comparing relativistic and nonrelativistic mirror-confined plasmas with the same ratio of confining potential to electron temperature $|e\phi|/T_e$ and the same mirror ratio, the net result of relativistic effects is a decrease in the confinement time by a factor of $S\equiv (1+15T_e/8m_ec^2)(1+2|e\phi|/m_ec^2)$.
In Sec.~\ref{sec:simulations}, we make use of finite-element simulations to verify the analytical estimates, finding good agreement between theory and simulation. 
In Sec.~\ref{sec:ambipolarPotential} we discuss the effect of this loss of confinement on the ambipolar potentials, before concluding in Sec.~\ref{sec:discussion} with a forward-looking discussion on how other effects, such as radiation, might also affect the results.


\section{Diffusion Equation} \label{sec:diffusionEquation}

We begin in this section with a derivation of the relativistic diffusion equation in appropriate coordinates.
The collisions of hot, relativistic particles with a largely nonrelativistic bulk population can be modeled as a momentum-space Fokker-Planck equation:\cite{Matsuda1986RelativisticMultiregion,Mosher1975InteractionsRelativistic,Braams1987DifferentialForm,Bernstein1981RelativisticTheorya,Karney1985EfficiencyCurrent}
\begin{align}
    \pa{f_a}{t} &= \pa{}{\ve{p}} \cdot \left[ \ve{A}_a f_a + \ve{D}_a \cdot \pa{f_a}{\ve{p}} \right],
\end{align}
where
\begin{align}
    \ve{D}_a &= \sum_b C_{ab} \left\{  Y_{ab} \frac{\gamma_a}{2}  \frac{p^2 \ve{I} - \ve{p}\ve{p}}{p^3}  + \gamma_a^3 \frac{m_a^2 T_b \ve{p}\ve{p}}{m_b p^5} \right\},
\end{align}
\begin{align}  
    \ve{A}_a &= \sum_b C_{ab} \left\{ \gamma_a^2 \frac{m_a \ve{p} }{m_b p^3} \right\},\\
    C_{ab} &= 4\pi n_b m_a Z_a^2 Z_b^2 e^4 \lambda_{ab},\\
    Y_{ab} &= 1 - \frac{m_a^2 T_b}{m_b p^2}, \\
    \gamma_a &= \sqrt{1 + \frac{p^2}{m_a^2 c^2}}.
\end{align}
Here, $e$ is the elementary charge, $c$ is the speed of light, $Z_a$, $m_a$, and $n_a$ are the charge, mass, and density of species $a$, and $\lambda_{ab}$ is the Coulomb logarithm.
All quantities are in Gaussian units.

This equation can be transformed to more useful coordinates by using the coordinate-invariant form of the diffusion equation:
\begin{align}
    \sqrt{g} \pa{f_a}{t} &= \pa{}{\ve{x}} \cdot \left[ \sqrt{g} \left(\ve{A}_a f_a + \ve{D}_a \cdot \pa{f_a}{\ve{x}}\right) \right],
\end{align}
where, using summation notation,
\begin{align}
    g_{ij} &= \pa{p^m}{x^i}\pa{p^n}{x^j} g_{mn},\\
    D_a^{ij} &= \pa{x^i}{p^m}\pa{x^j}{p^n} D^{mn},\\
    A_a^{i} &= \pa{x^i}{p^m} A^{m}.
\end{align}

It is common in the mirror literature to assume gyrotropy and use the coordinates:
\begin{align}
    \bar{x} &= v/v_{\text{th},a};\qquad v_{\text{th},a} \equiv \sqrt{2 T_a / m_a};\\
    \bar{\xi} &= v_\parallel/v.
\end{align}
The natural relativistic generalization of these coordinates is:
\begin{align}
    x &= p/p_{\text{th},a}; \qquad p_{\text{th},a} \equiv \sqrt{2 m_a T_a};\\
    \xi &= p_\parallel/p.
\end{align}

Performing the coordinate transformation and dropping the subscripts for species $a$, we find:
\begin{align}
    \tau_{0} \pa{f}{t} &= \frac{1}{x^2} \pa{}{x} \left(\gamma^2 Z_\parallel f + \frac{\gamma^3}{2 x} \pa{f}{x}\right) \notag \\
    &\hspace{0.1in}+ \frac{\gamma}{x^3} \lp Z_\perp - \frac{1}{4 x^2} \rp \pa{}{\xi} \left[(1-\xi^2) \pa{f}{\xi} \right], \label{eq:diffusionEquation}
\end{align}
where:
\begin{align}
    Z_\parallel &= \frac{\sum_b n_b Z_b^2 \lambda_{ab} / m_b}{\sum_b n_b Z_b^2 \lambda_{ab} T_b/ m_b T_a},\\
    Z_\perp &= \frac{1}{2} \frac{\sum_b n_b Z_b^2 \lambda_{ab} }{\sum_b n_b Z_b^2 \lambda_{ab} m_a T_b / m_b T_a},
\end{align}
and the Lorentz factor is given by:
\begin{align}
    \gamma &= \sqrt{1 + 2 \chi x^2}, \label{eq:gammaNondim}
\end{align}
which depends on the critical new parameter:
\begin{align}
    \chi &= T_a / m_a c^2.
\end{align}
Often in the literature, the parameter $\chi$ is referred to as $\theta_T$; we use $\chi$ to avoid confusion with the angular coordinate $\theta$.

Finally, the collisional timescale $\tau_0$ is given by:
\begin{align}
	\tau_{0}^{-1} &= 4\pi e^4 \sum_b \frac{n_b Z_a^2 Z_b^2 \lambda_{ab} m_a^2}{m_b p_{\text{th},a}^3} \frac{T_b}{T_a}.
\end{align}
As typical examples, $\tau_0 = 400$ $\mu$s in a 20 keV DT fusion plasma at $n_i = 10^{14}$ cm$^{-3}$, and $\tau_0 = 5$ ms in a 150 keV electron, 300 keV ion pB11 fusion plasma at $n_i = 10^{14}$ cm$^{-3}$.

The error terms corrections in Eq.~(\ref{eq:diffusionEquation}) are of the order of the bulk relativistic parameter $\mathcal{O}(\chi_s)$, which will be negligible for the ions but finite for the electrons.
Thus, effects of a relativistic electron bulk $\chi$ only substantially impact the parallel diffusion, not the perpendicular diffusion.
Importantly, the finite-$\chi_e$ modifications also do not affect the thermodynamic steady state of the parallel diffusion equation, which is given by the Einstein relation.
Thus, as we will see in Sec.~\ref{sec:analyticalEstimates}, the neglected error terms should have only an extremely mild impact (less than $\mathcal{O}(\chi_s)$) on the results.

In the nonrelativistic limit $\chi \rightarrow 0$, Eq.~(\ref{eq:diffusionEquation}) reduces almost to the diffusion equation in Ref.~\onlinecite{Najmabadi1984}.
The difference comes from the fact that Ref.~\onlinecite{Najmabadi1984} had $Z_\parallel \rightarrow 1$.
When $Z_\parallel = 1$, then Eq.~(\ref{eq:diffusionEquation}) says that species $a$ approaches a Maxwell-J\"uttner distribution with temperature $T_a$.
Thus, Najmabadi's equation assumes that the temperature of species $a$ (as used in the normalization for $x$) is consistent with the temperature it is driven to by collisions with all species in the plasma.
This is often a safe assumption, especially when the losses due to deconfinement occur far out on the tail, but it is good to state explicitly. 
For the rest of the paper, we take $Z_\parallel = 1$.

\section{Relativistic Trapping Condition} \label{sec:lossCone}

In addition to the diffusion equation, the trapping condition for particles inside a magnetic mirror is modified if one considers the effects of relativity.\cite{Matsuda1986RelativisticMultiregion} 
In this section we review this modified loss cone boundary, and transform it to the dimensionless mirror coordinates.

Consider a particle in a mirror with a magnetic field $B_0$ and (species-dependent) potential energy $U_{a0}$ at the midplane, and corresponding quantities $B_1$ and $U_{a1}$ at the mirror boundary.
As the particle traverses from the midplane to the boundary, there will be two conserved invariants: the relativistic energy:
\begin{align}
    \epsilon = \sqrt{m_a^2 c^4 + (p_\parallel^2 + p_\perp^2) c^2} + U_a, \label{eq:relEnergy}
\end{align}
and the relativistic magnetic moment:\cite{Northrop1966,Oskoui2014,vonderLinden2021ConfinementRelativistic}
\begin{equation}
\mu =\frac{p_{\perp}^{2}}{2m_a B}.\label{eq:relMu}
\end{equation}

The boundary between trapped and passing particles, known as the loss cone, is the point at which $p_\parallel  = 0$ at the mirror boundary.
Using the definitions and invariance of $\epsilon$ and $\mu$,
as well as the normalizations from Sec.~\ref{sec:diffusionEquation}, this boundary can be written as:
\begin{align}
    R(1-\xi^2) &= 1- \frac{u}{x^2} \left(\gamma - \frac{1}{2} \chi u \right), \label{eq:lossConeRelativistic}
\end{align}
where we have defined the mirror ratio $R \equiv B_1/B_0$, and the normalized potential $u = (U_{a1}-U_{a0})/T_a$.
Note that this reduces to the nonrelativistic expression [Eq.~(9) of Ref.~\onlinecite{Najmabadi1984}] when $\chi \rightarrow 0$.
The error terms corrections in Eq.~(\ref{eq:diffusionEquation}) are of the order of the bulk relativistic parameter $\mathcal{O}(\chi)$, which we will take to be small even when the tail electrons are highly relativistic.

\begin{figure}[t]
	\centering
	\includegraphics[width=\linewidth]{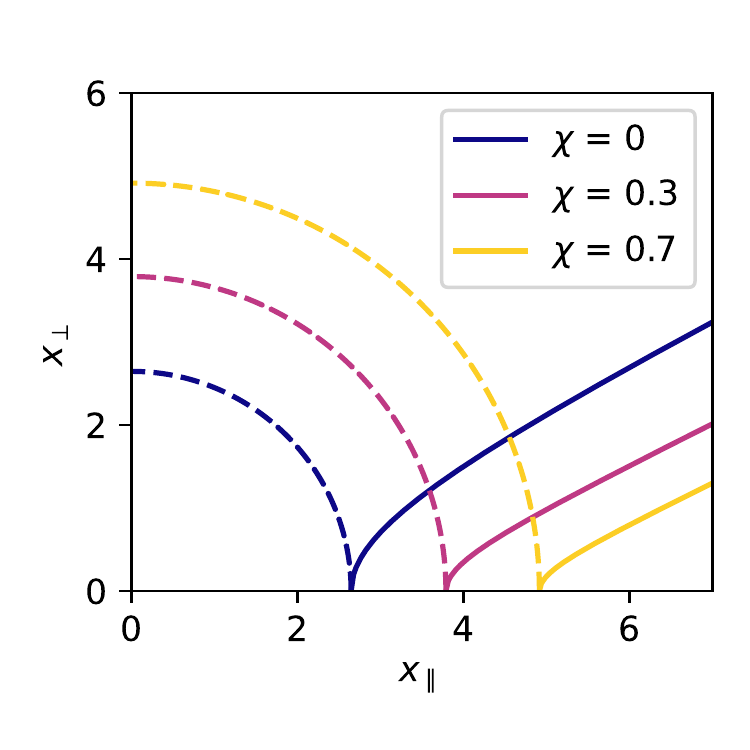}
	\caption{Loss cone (solid lines) in normalized perpendicular and parallel energy space for several values of the temperature-to-rest-mass ratio $\chi$ for $u=7$, $R=5$.
		While it appears that the loss cone vertex moves further out on the high-energy tail as the plasma becomes more relativistic, the energetic accessibility remains the same in each case.
		This is shown in the figure by the three dotted lines, which each represent the same Boltzmann factor $(\gamma-1)/\chi = u$ at the different values of $\chi$.}
	\label{fig:lossConeVsChiU}
\end{figure}

It is common to plot the loss cone in terms of the parallel and perpendicular dimensionless momenta $x_\parallel = x \xi$ and $x_\perp = x \sqrt{1-\xi^2}$.
Such a plot is shown in Fig.~\ref{fig:lossConeVsChiU} for $R = 5$ and $u=7$.
It is important to note that, while it looks like the vertex of the loss cone is moving ``further out on the tail'' of high $x$, the vertex is in fact equally energetically accessible in each case.
This can be seen by noting that the temperature-normalized escape energy (and thus the Boltzmann factor) is the same in each case, equal to the normalized rest energy plus the normalized confining potential energy, regardless of how relativistic the plasma is.

\section{Analytical Estimates} \label{sec:analyticalEstimates}

A crude estimate of the confinement time, first proposed in Ref.~\onlinecite{Kaufman1956}, can be made by assuming that the particles arrange themselves close to a Maxwell-J\"uttner distribution, and that the fraction $f_L$ with sufficient energy then scatter into the loss cone on the perpendicular diffusion timescale $\tau_\perp$.
This gives a confinement time estimate of:
\begin{align}
    \tau_C \sim \frac{\tau_\perp}{f_L}, \label{eq:confinementCrudeEstimate}
\end{align}
where 
\begin{align}
    \tau_\perp &= \frac{x_c^3}{Z_\perp \gamma_c} \tau_0, \label{eq:tauPerpRel}\\
     f_L &\equiv\int_{x_c}^\infty 4\pi x^2 f_\text{MJ}(x) dx. \label{eq:fL}
\end{align}
Here, the Maxwell-J\"uttner distribution is given in normalized coordinates by:
\begin{align}
    f_\text{MJ}(x)=\frac{1}{\sqrt{2} \pi} \frac{\chi^{1/2}}{K_2(1/\chi)} e^{-\gamma/\chi}. \label{eq:fMJNondim}
\end{align}
Here, $x_c$ is the minimum possible normalized momentum that can escape into the loss cone through perpendicular scattering, with $\gamma_c =\sqrt{1+2\chi x_c^2}$ its associated Lorentz factor, and $K_n(z)$ is the modified Bessel function of the second kind.
We can calculate $x_c$ either from solving Eq.~(\ref{eq:lossConeRelativistic}) for $x$ at $\xi = 1$, or by taking equating the midplane particle energy with the energy needed to escape the potential at $p_\perp = 0$ (i.e., taking $\epsilon = m_a c^2 + U_{a1}-U_{a0}$ at $p_\perp = 0$) and normalizing.
Either way, we find:
\begin{align}
    x_c &= \sqrt{u + \frac{1}{2} \chi u^2}, \label{eq:xc}\\
    \gamma_c &= 1 + \chi u. \label{eq:gammac}
\end{align}

\subsection{Approximate Forms of the Integrals}

The Maxwell-J\"uttner integral, although not exactly expressible in closed form, can be approximated in both the nonrelativistic and highly relativistic limits.
If $\chi \sim \chi u \ll 1$, i.e.,~if the energy of the confining potential is much less than the rest mass, then we can expand the Bessel function in large argument:
\begin{align}
    K_2\left(1/\chi\right) \approx \sqrt{\frac{\pi}{2}} \chi^{1/2} e^{-1/\chi},
\end{align}
so that
\begin{align}
    f_\text{MJ}(x)\approx \frac{1}{\pi^{3/2}} e^{-(\gamma-1)/\chi}.
\end{align}
We can also Taylor expand $\gamma$ in small $\chi x^2$, yielding the Maxwell-Boltzmann distribution:
\begin{align}
    f_\text{MJ}(x)\approx \frac{1}{\pi^{3/2}} e^{-x^2}; \quad \chi u \ll 1. \label{eq:fMJNonrel}
\end{align}
In this limit, we can also replace the lower integral bound by $x_c \approx \sqrt{u}$.
In the limit $u \gg 1$, the integral then evaluates to:
\begin{align}
    f_L \approx \frac{2}{\sqrt{\pi}} u^{1/2} e^{-u}; \quad \chi u \ll 1 \text{ and } u \gg 1. \label{eq:fLNonrel}
\end{align}

Alternatively, if $\chi u \gg 1$, i.e.,~if the energy of the confining potential is much greater than the rest mass, then it makes sense to perform the integral over $\gamma$ rather than $\chi$, from the lower bound at $\gamma_c = 1+\chi u$:
\begin{align}    f_{L}&=\int_{\gamma_{c}}^{\infty}\frac{\gamma\sqrt{\gamma^{2}-1}e^{-\frac{\gamma}{\chi}}}{\chi K_{2}\left(1/\chi \right)}d\gamma\\
&\approx\int_{\gamma_{c}}^{\infty}\frac{(\gamma^2-\tfrac{1}{2})e^{-\frac{\gamma}{\chi}}}{\chi K_{2}\left(1/\chi \right)}d\gamma\\
&=\frac{ \left[1+\chi \left(u+1\right)\right]^{2}+\chi^{2}-\frac{1}{2}}{K_{2}\left(1/\chi\right)}e^{-u-\frac{1}{\chi}}\label{eq:fLUltrarel}\\
&\approx \frac{\chi^2 u^2}{K_2(1/\chi)} e^{-u-1/\chi} \label{eq:fLUltrarel2}.
\end{align}

\begin{figure}
    \centering
    \includegraphics[width=\linewidth]{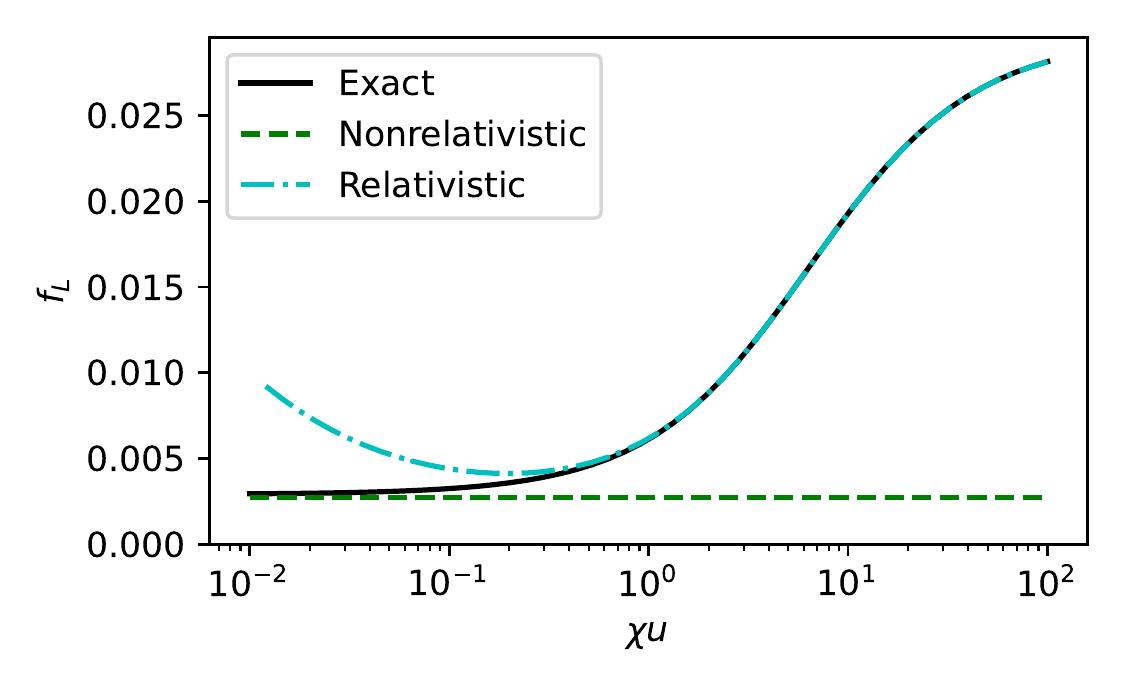}
    \caption{Fraction of particles above the loss energy for a Maxwell-J\"uttner distribution for $u=7$, as a function of relativistic parameter $\chi u$.
    Shown are the exact solution (black solid), the approximate nonrelativistic formula from Eq.~(\ref{eq:fLNonrel}) (green dashed), and the approximate relativistic form from Eq.~(\ref{eq:fLUltrarel}) (cyan dash-dotted).} 
    \label{fig:integralApproximations}
\end{figure}

The approximate formulae from Eqs.~(\ref{eq:fLNonrel}) and (\ref{eq:fLUltrarel}) are plotted in Fig.~\ref{fig:integralApproximations} for $u=7$ as a function of $\chi u$.
The agreement is quite good, except at marginally relativistic values.

\subsection{Refining the Approximation Using Existing Solutions}
It is well known that in addition to the above factors, there is a $\sim \log R$ dependence of the confinement time on the mirror ratio.
To get these factors, we can make use of existing solutions from the literature. 
Specifically, we make use of the solution in Ref.~\onlinecite{Najmabadi1984}:
\begin{align}
    \frac{\tau_{C,\chi = 0}}{\tau_0} =  \frac{1}{Z_\perp I} \frac{\sqrt{\pi}}{4} u_\text{eff} e^{u_\text{eff}} \left[\log \left(\frac{w+1}{w-1}\right) - 0.84 \right],  \label{eq:tauCNonrel}
\end{align}
where
\begin{align}
    w^2 &\equiv 1+\frac{1}{R \lp Z_\perp - \frac{1}{4u_\text{eff}} \rp},\\
    I &= -\frac{1}{4Z_\perp} + \left(1+\frac{1}{4Z_\perp}\right) u_\text{eff} e^{u_\text{eff}} E_1(u_\text{eff}),
\end{align}
$E_1(y) = \int_y^\infty \tfrac{e^{-t}}{t} dt$ is the exponential integral function, and and $u_\text{eff} = u+\log w$.
Note that $w$ is defined incorrectly the second time it appears in Ref.~\onlinecite{Najmabadi1984}, a typo which unfortunately made it into the standard review.\cite{Post1987}
Note also that we have defined $I$ slightly differently here, so that it is approximately 1 as $u_\text{eff} \rightarrow \infty$, to make the scaling with $Z_\perp$ more explicit.

To extend from this solution, we can multiply by the ratio between the relativistic and nonrelativistic formulas, i.e.,
\begin{align}
    \frac{\tau_C}{\tau_{C,\chi = 0}} \approx  \frac{f_{L, \chi = 0}}{f_L} \frac{\tau_\perp}{\tau_{\perp,\chi = 0}}. \label{eq:tauCThFull}
\end{align}

Obviously, as $\chi \rightarrow 0$, this ratio goes to one.
To find this ratio in the relativistic-tail ($\chi u \gg 1$), nonrelativitic-bulk ($\chi \ll 1$) limit, we can plug in Eq.~(\ref{eq:fLUltrarel2}) and expand the Bessel function in small $\chi$, yielding:
\begin{align}
    \frac{\tau_{C,\chi u \gg 1}}{\tau_{C,\chi u = 0}} \sim \frac{1}{2} \frac{1+\tfrac{15}{8} \chi}{\chi u}. 
\end{align}
This can be made into a formula that agrees with both the nonrelativistic and ultrarelativistic limits via asymptotic matching:
\begin{align}
    \frac{\tau_{C}}{\tau_{C,\chi = 0}} \approx S(\chi,u) \equiv \frac{1+\tfrac{15}{8} \chi}{1+2 \chi u} . \label{eq:netEffectRelativityScaling}
\end{align}

Importantly, for pB11 fusion, with $T_e \approx 150$ keV and thus $\chi \approx 0.3$, Eq.~(\ref{eq:netEffectRelativityScaling}) shows that the change in confinement time due to relativistic effects is large.
For a normalized confining potential of $u = 5$, the confinement time is reduced by a factor of 2.5 from Ref.~\onlinecite{Najmabadi1984}'s nonrelativistic results, whereas at $u = 10$, it is reduced by a factor of 5.

\section{Verification of Confinement Times} \label{sec:simulations}

To verify the confinement time estimates from Sec.~\ref{sec:analyticalEstimates}, we perform numerical Fokker-Planck simulations for the diffusion process in Eq.~(\ref{eq:diffusionEquation}), with the loss cone described by Eq.~(\ref{eq:lossConeRelativistic}).
Without loss of generality, we take $\tau_0 = 1$.
To maintain the resolution of the loss cone, we work in the coordinates $(x,\theta)$, where $\cos \theta \equiv \xi$.
In these coordinates, the metric is given by $\sqrt{g} = 4\pi x^2 \sin \theta$.
To avoid problems at the boundaries, we modify the denominator of the diffusion and advection operators so that they do not diverge as $x \rightarrow 0$.
Since we are only interested in the steady-state process, we add a particle source $s(x,\theta)$, and then solve the steady-state diffusion equation, which (with all the above modifications) is given by:
\begin{align}
	0 &=  \pa{}{x} \left[\sqrt{g}\left( \frac{\gamma^2 x}{x^3+x_0^3} f + \frac{\gamma^3}{2 (x^3+x_0^3)} \pa{f}{x}\right)\right] \notag \\
	&\hspace{0.05in}+  \pa{}{\theta} \left[\sqrt{g} \left(\frac{\gamma }{x^3+x_0^3} \lp Z_\perp - \frac{Z_\perp}{1 + c_0  + 4 Z_\perp x^2} \rp \pa{f}{\theta} \right)\right] \notag\\
	&\hspace{0.05in} + \sqrt{g} s(x,\theta)  ;\label{eq:diffusionEquationSim}\\
	\sqrt{g} &= 4\pi x^2 \sin \theta.
\end{align}
The relevant boundary conditions are reflecting everywhere except at the loss cone, where the solution must go to 0.
Once the solution is obtained subject to the boundary conditions, the confinement time is then given by the ratio between the integrated density and integrated source:
\begin{align}
    \bar{\tau}_C \equiv \frac{\tau_C}{\tau_0} = \frac{\int f \sqrt{g} dx d\theta}{\int s \sqrt{g} dx d\theta}. \label{eq:tauCSim}
\end{align}

To actually solve Eq.~(\ref{eq:diffusionEquationSim}) subject to the rather complex boundary conditions, we use the DolfinX finite-element library\cite{Scroggs2022ConstructionArbitrary} with a mesh created using the gmsh library.\cite{Geuzaine2009Gmsh3D}
Since the domain is theoretically infinite in $x$, we must choose a maximum value $x_\text{max}$ to define the upper edge of the domain. 
We choose this maximum to correspond to $K$ $e$-foldings of the Maxwell-J\"uttner distribution, which means that for a given $u$, $x_\text{max} = x_c |_{(u\rightarrow u+K)}$, i.e., we replace $u$ with $u+K$ in Eq.~(\ref{eq:xc}) to determine the large-$x$ edge of the domain.

We also must choose a specific source function.
We choose a form that smoothly goes to zero on the boundaries, i.e.,
\begin{align}
    s = x^2 \theta^2 \left(\tfrac{\pi}{2}-\theta \right)^2 e^{-x^2/x_{s0}^2} .
\end{align}
For the simulations presented here, these non-physical parameters are $K = 7$, and $x_0 = x_{s0} = 0.1$ and $c_0 = 0.2$.

\begin{figure}
	\centering
	\includegraphics[width=\linewidth]{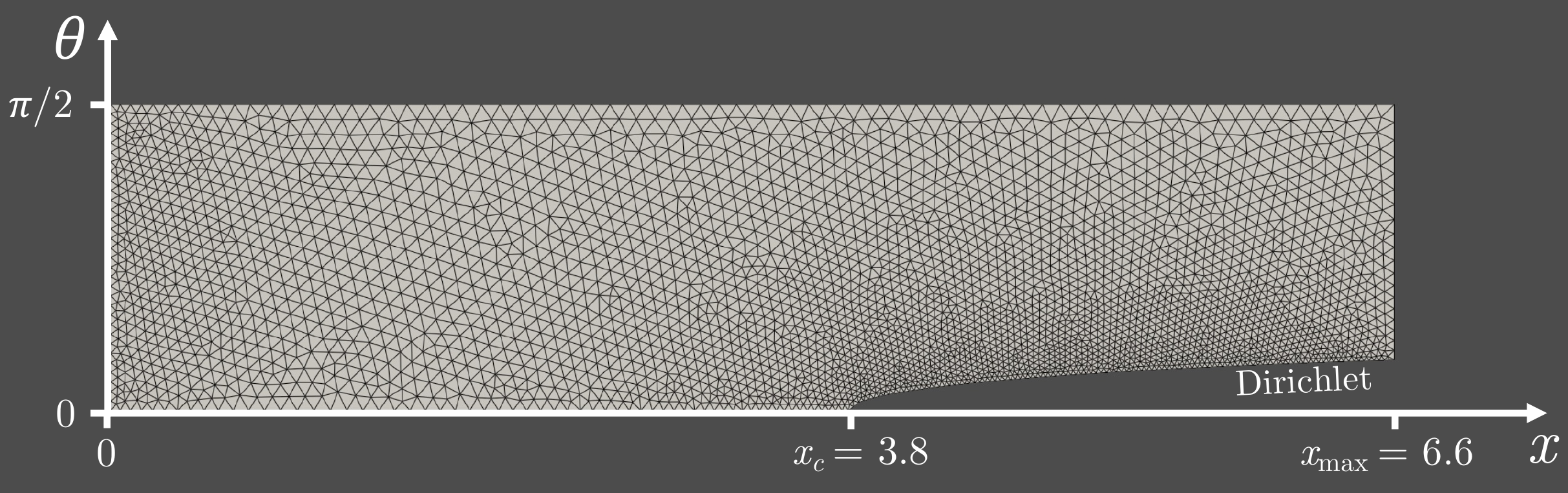}
	\caption{Example gmsh-generated mesh for $u=7$, $R=5$, and $\chi = 0.3$, with a mesh size $\Delta x = \Delta \theta = 0.1$.
		The mesh is chosen to be slightly more refined near the source at $x=0$, and the loss cone.
		Dirichlet conditions ($f=0$) are enforced at the loss cone, and zero-flux conditions at all other boundaries.}
	\label{fig:mesh}
\end{figure}

\begin{figure}
	\centering
	\includegraphics[width=\linewidth]{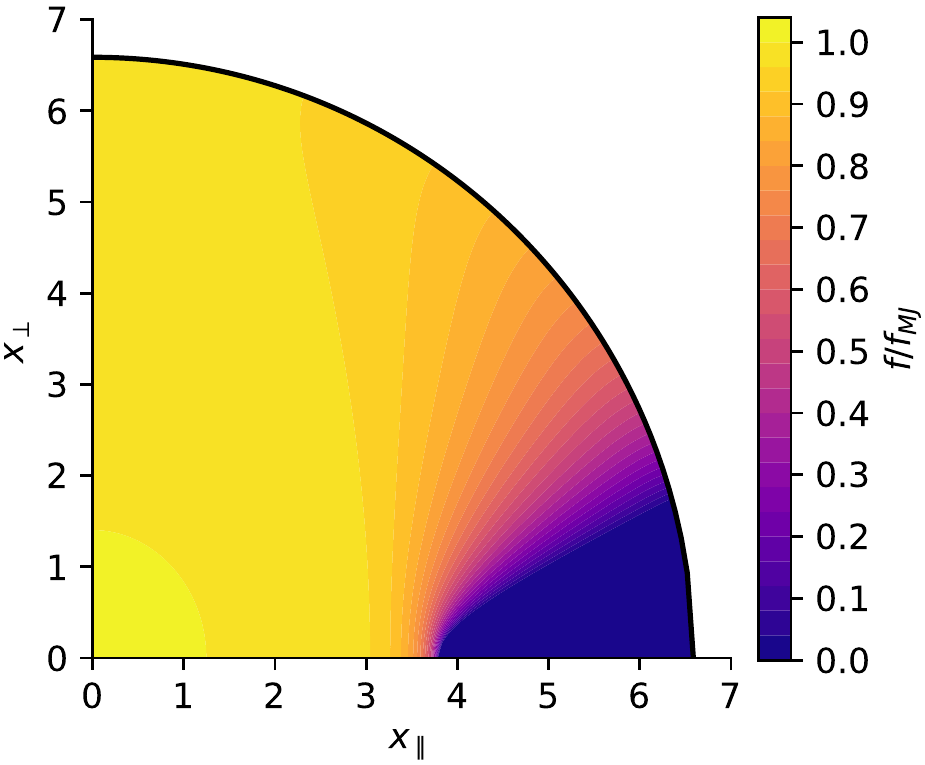}
	\caption{Numerical solution to the the relativistic Fokker-Planck diffusion equation [Eq.~(\ref{eq:diffusionEquationSim})] on the mesh in Fig.~\ref{fig:mesh} for $u=7$, $R=5$, $\chi = 0.3$, and $Z_\perp = 1$.
		The solution is plotted relative to the dimensionless Maxwell-J\"uttner distribution from Eq.~(\ref{eq:fMJNondim}).
		We can see that the solution is very close to $f_\text{MJ}$ except in the immediate region of the loss cone, which is why the estimate in Eq.~(\ref{eq:tauCThFull}) is fairly accurate.}
	\label{fig:exampleSolution}
\end{figure}

We use a mesh size of of $\Delta x = \Delta \theta = 0.1$, with double resolution near the loss cone and low-energy source boundary, and third-order continuous Galerkin finite elements.
Pseudo-convergence testing suggests that this combination results in a relative error in the integrated quantities of less than 1\%, while initializing and running in less than a second.
An example mesh is shown in Fig.~\ref{fig:mesh}, for $u = 7$, $R=5$, and $\chi = 0.3$.
The solution to Eq.~(\ref{eq:diffusionEquationSim}) on this mesh for $Z_\perp = 1$ is shown in Fig.~\ref{fig:exampleSolution}.
The solution is quite close to a Maxwell-J\"uttner distribution, except in the immediate vicinity of the loss cone, which justifies the approximation method from Eq.~(\ref{eq:tauCThFull}).

\begin{figure}
    \centering
    \includegraphics[width=\linewidth]{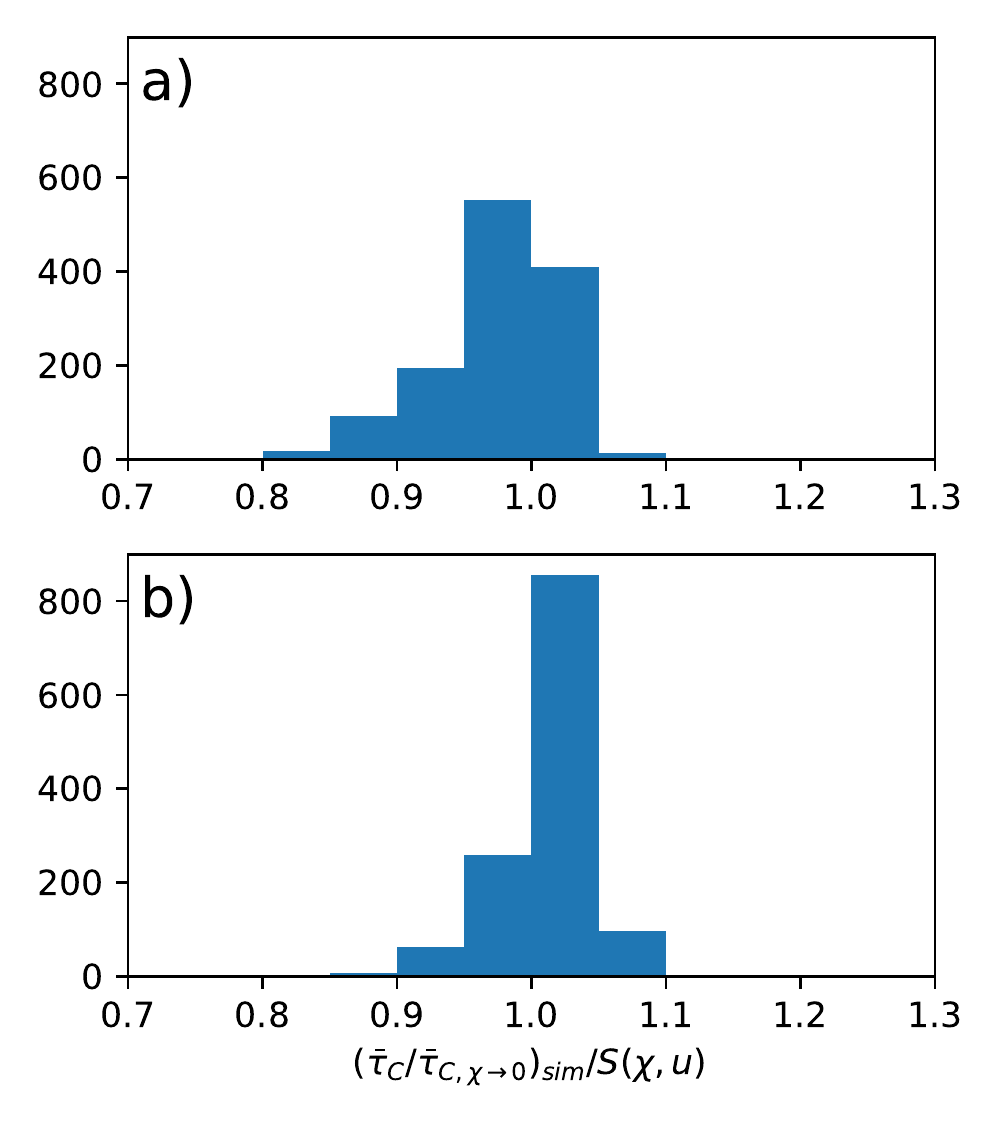}
    \caption{Histogram comparing the ratio of simulated relativistic to nonrelativistic ($\chi = 10^{-4}$) confinement time, as calculated from Eq.~(\ref{eq:tauCSim}), to (a) the exact prediction from Eq.~(\ref{eq:tauCThFull}), using Eqs.~(\ref{eq:tauPerpRel}-\ref{eq:fL}), and (b) the asymptotic scaling prediction $S(\chi,u)$ from Eq.~(\ref{eq:netEffectRelativityScaling}), for all simulations in the dataset.
    Surprisingly, the asymptotic approximation outperforms the more accurate calculation, agreeing within 10\% in nearly all cases.}
    \label{fig:tauCVsScaling}
\end{figure}

\begin{figure}
    \centering
    \includegraphics[width=\linewidth]{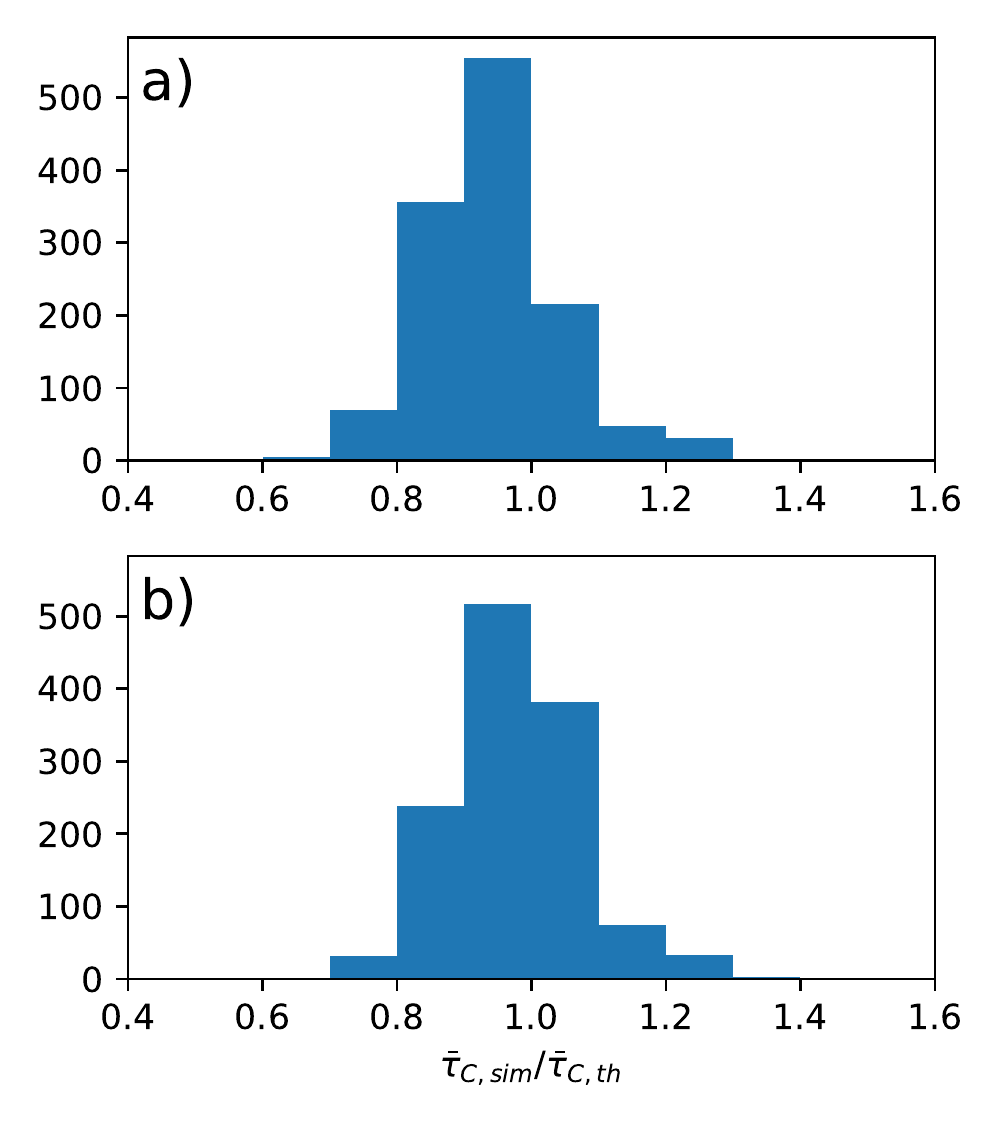}
    \caption{Histogram of the ratio between the simulated confinement time [Eq.~(\ref{eq:tauCSim})] vs. the theoretical confinement time [Eqs.~(\ref{eq:tauCNonrel}) and (\ref{eq:tauCThFull})], using (a) the full numerical solution of the factor in Eq.~(\ref{eq:tauCThFull}) from Eqs.~(\ref{eq:tauPerpRel}-\ref{eq:fL}), and (b) the asymptotic approximation from Eq.~(\ref{eq:netEffectRelativityScaling}).
    The theoretical form almost always falls within a factor of 30\% of the simulated value.}
    \label{fig:tauCVsTauCTh}
\end{figure}

To test the accuracy of the estimate for $\tau_C$ in Eq.~(\ref{eq:tauCThFull}), we performed a parameter scan for all 1280 combinations of the parameters:
\begin{align}
    u &\in \{4,6,8,10,12,14,16,18\},\\
    R &\in \{5,10,15,20\},\\
    \chi &\in \{0.0001,0.001,0.01,0.03,0.1,0.3,0.5,1\},\\
    Z_\perp &\in \{0.5,1,3,5,10\}.
\end{align}
In Fig.~\ref{fig:tauCVsScaling}, we see that the ratio of $\tau_C / \tau_{C,\chi\rightarrow 0}$ closely matches the theoretical result both in terms of the full formula implied by Eqs.~(\ref{eq:tauCThFull}) and (\ref{eq:tauPerpRel}-\ref{eq:fL}), and the asymptotic estimate from Eq.~(\ref{eq:netEffectRelativityScaling}).
In fact, the asymptotic formula actually performs better, matching to within 10\% almost the entire dataset.

In Fig.~\ref{fig:tauCVsTauCTh}, we see that combined with the existing nonrelativistic formula [Eq.~(\ref{eq:tauCNonrel})] from Ref.~\onlinecite{Najmabadi1984}, the relativistic theory estimates the confinement time quite well, with an error usually substantially less than 30\%, whether using the full theory or asymptotic scaling formula.
It should be emphasized that this solution involved no fitted parameters, but just an informed extension of existing analytical formulas.

\begin{figure}
    \centering
    \includegraphics[width=\linewidth]{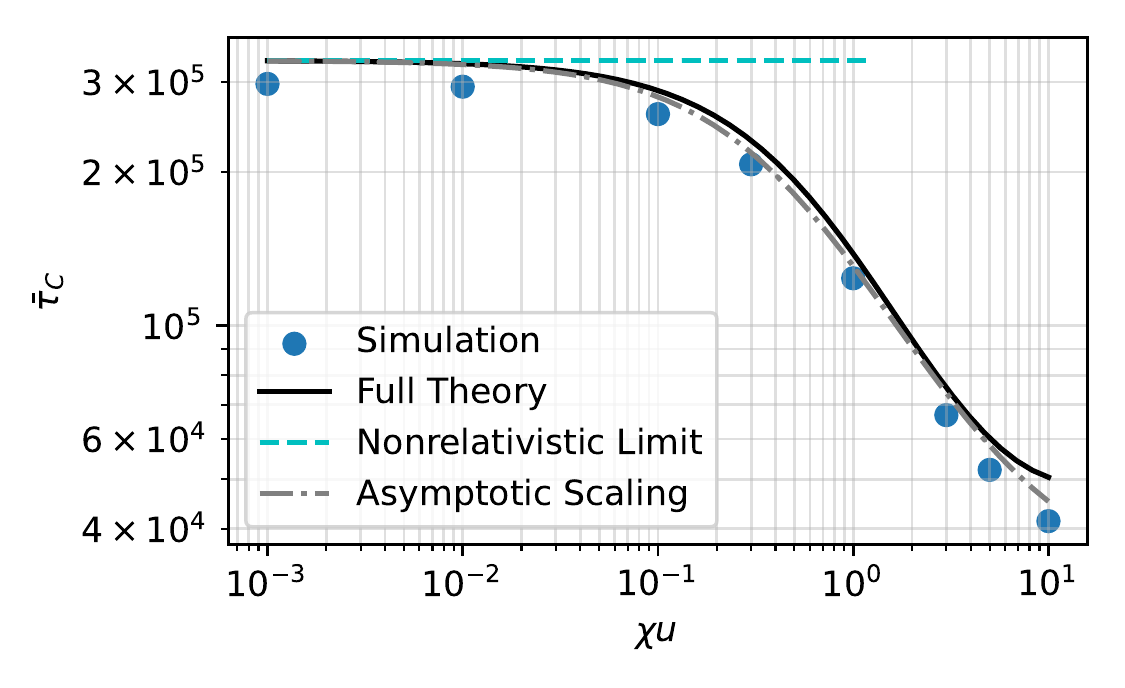}
    \caption{Dimensionless confinement time vs relativistic factor $\chi u$ for $u=10$, $R=10$, and $Z_\perp = 1$, comparing numerical simulations to the full theory [Eqs.~(\ref{eq:tauCThFull}) and (\ref{eq:tauCNonrel})], nonrelativistic limit [Eq.~(\ref{eq:tauCNonrel})], and asymptotic scaling formula [Eqs.~(\ref{eq:netEffectRelativityScaling}) and (\ref{eq:tauCNonrel})].
    The theory matches well.}
    \label{fig:tauCVsChiU}
\end{figure}

Finally, for a given value of $u$, $R$, and $Z_\perp$, we can look at the confinement time as a function of $\chi u$.
Such a scan is shown in Fig.~\ref{fig:tauCVsChiU} for $u=10$, $R = 10$, and $Z_\perp=1$.
We see that both the full theory and asymptotic scaling formula accurately capture the effects of the increasing relativistic parameter.
Interestingly, the overperformance of the asymptotic scaling relative to the full formula seems to occur near $\chi = 1$, where its validity begins to break down.
To understand why the simple scaling seems so robust, one would have to perform a more complex analysis, approximately solving the relativistic diffusion problem analytically using a Pastukhov-style matching to the relativistic loss cone.
Of course, at the point of interest, the nonrelativistic-bulk approximation in the collision model also breaks down, so, while intriguing, this quirk is likely of little physical significance.
Instead, what is important is that the very simple asymptotic formula of Eq.~(\ref{eq:netEffectRelativityScaling}) captures the relativistic effects on the confinement time quite accurately.


\section{Effect on Ambipolar Potentials} \label{sec:ambipolarPotential}

The ambipolar potential results from ensuring that, when the loss rate of all species is taken into account, no net charge leaves the system.
In terms of the confinement time, this can be written:
\begin{align}
    \sum_s Z_s n_s \tau_{Cs}^{-1} = 0. \label{eq:ambipolarCondition}
\end{align}

Consider a plasma confined in a centrifugal mirror trap.
Each species $s$ will feel a confining potential:
\begin{align}
    U_s = Z_s e \phi + U_{C,s}, 
\end{align}
where $\phi$ is the electrostatic potential and $U_{C,s}$ is the species-dependent centrifugal potential.
Thus, there will be a corresponding species-dependent dimensionless potential:
\begin{align}
    u_s = \frac{Z_s e \phi}{T_s} + \frac{U_{C,s}}{T_s}.
\end{align}
Consider a rapidly-rotating plasma, with protons $p$, electrons $e$, and (optionally) a heavy species that is much better confined (due to the rotation) than the other two. 
Such a plasma describes, for example, a pB11 fusion plasma.
Using Eq.~(\ref{eq:tauCNonrel}) and the scaling relation Eq.~(\ref{eq:netEffectRelativityScaling}) in Eq.~(\ref{eq:ambipolarCondition}), we find an equation for the ambipolar potential of the form:
\begin{align}
    \bar{\phi} &= \frac{T_p}{T_p + T_e} \biggl[ u_{C,p} + \log \lp \frac{n_e}{n_p} \frac{\tau_{Dp}}{\tau_{De}} \rp - \log S\left(\chi,\bar{\phi}\right)\notag\\
    &\hspace{0.5in} + \log \left(\frac{u_{C,p} - \tfrac{T_e}{T_p} \bar{\phi}}{\bar{\phi}} \right) \biggr], \label{eq:ambipolarPhiImplicit}
\end{align}
where
\begin{align}
    \bar{\phi} &\equiv e \phi / T_e,\\
    u_{C,p} &\equiv U_{C,p} / T_p,\\
    \tau_{Ds} &\equiv \frac{\tau_{0s}}{Z_{\perp,s} } \left[\log\left(\tfrac{w_s+1}{w_s-1}\right)-0.84\right].
\end{align}

Equation~(\ref{eq:ambipolarPhiImplicit}) can be approximately solved in orders, by first ignoring the $\bar{\phi}$-dependent logarithms on the right hand side, and then adding in this logarithm as a correction based on the solution.
Since the equation can be typically be solved in this way with good accuracy, we see that the net effect of the relativistic corrections is to increase the ambipolar potential by an amount:
\begin{align}
    \Delta \bar{\phi} \approx \frac{T_i}{T_e + T_i} \log \left( \frac{1 + 2\chi \bar{\phi}}{1 + \tfrac{15}{8} \chi} \right).
\end{align}

\section{Discussion and Conclusion} \label{sec:discussion}

In this paper, we have generalized the nonrelativistic work of Refs.~\onlinecite{Pastukhov1974,Cohen1978,Cohen1980,Najmabadi1984,Post1987} to incorporate relativistic corrections to the confinement time of particles in mirror, assuming that the bulk plasma is nonrelativistic.
Both theory and simulations showed that the effect of relativity is to decrease the electron confinement time, and thus increase the ambipolar potential.
Such effects are important to take into account in extrapolating to the extreme high-temperature plasmas ($T_e \sim 160$ keV, $\chi \sim 0.3$) necessary for aneutronic fusion. \cite{Hay2015Ignition,Putvinski2019,kolmes2022waveSupported,Ochs2022ImprovingFeasibility}

However, as the plasma grows more relativistic, radiative effects tend to become more important as well, with both bremsstrahlung and synchrotron radiation\cite{Mlodik2023SensitivitySynchrotron} becoming an important part of the energy balance.
These processes are often modeled in Fokker-Planck form in the study of runaway electrons in tokamaks.\citep{Stahl2015,Breizman2019}
Of the two of these, synchrotron radiation is usually more powerful.
While a detailed analysis of the impact of radiation is outside the scope of this paper, here we estimate when this term is likely to be important.

When the plasma is optically thin to synchrotron radiation (which is often the case for high harmonics emitted by the hottest parts of the distribution), the emitted radiation leads to a pure drag term in the Fokker-Planck equation, with a characteristic timescale given by:\cite{Breizman2019}
\begin{align}
    \tau_S &= \frac{\tau_{S0}}{\gamma(1-\xi^2)},\\
    \tau_{S0} &= \frac{3}{2}\frac{m_{e}^{3}c^{5}}{e^{4}B^{2}}.
\end{align}
Meanwhile, the parallel collisional drag timescale is given from Eq.~(\ref{eq:diffusionEquation}) by:
\begin{align}
    \tau_{c\parallel} = \tau_0 \frac{x^3}{\gamma^2} .
\end{align}
Using the definitions of $x$ and $\tau_0$, we thus find at the loss-cone-relevant energy:
\begin{align}
    \frac{\tau_{c\parallel}}{\tau_{S}} &= \frac{2}{3 \lambda_e}\frac{\Omega_e^{2}}{\omega_{\textrm{pe}}^{2}}\frac{\left(\gamma_c^{2}-1\right)^{\frac{3}{2}}}{\gamma_c} \left(1-\xi^{2}\right)\\
    &= \frac{2^{5/2}}{3 \lambda_e}\frac{\Omega_e^{2}}{\omega_{\textrm{pe}}^{2}}(\chi u)^{3/2}\frac{\left(1 +\chi u/2 \right)^{\frac{3}{2}}}{1+\chi u} \left(1-\xi^{2}\right),
\end{align}
where  $\lambda_{e}$ is the Coulomb logarithm, while  $\Omega_e$ and $\omega_{pe}$ are the nonrelativistic election cyclotron and plasma frequencies, respectively.
Similar expressions appeared in Refs.~\onlinecite{McTiernan1990, Fisch1990} (although there is a typo in Ref.~\onlinecite{Fisch1990}, where $\lambda_e$ appears in the numerator instead of the denominator).
Note the angular dependence of this condition; the synchrotron radiation is maximized when most of the energy is in the perpendicular direction $\xi \sim 0$, while the loss cone is located in the parallel direction $\xi \sim 1$.
Thus, even when synchrotron radiation becomes important for the overall Fokker-Planck solution, its effects will first be felt far from the loss cone.
Note that a more general theory, that takes into account both the drag term due to radiation reaction and the diffusion term due to absorption of synchrotron radiation, is presented in Ref.~\onlinecite{Bornatici_FP1994}.

There are other deconfining effects as well that we have not here considered.
For instance, our model assumes that the magnetic field variation is sufficiently slow that loss of adiabaticity\cite{Saitoh2016ChaosEnergetic,vonderLinden2021ConfinementRelativistic} can be neglected relative to collisions.
In addition, we have neglected the deconfining effects of cold particle flows from the mirror throat. \cite{Konkashbaev1978PossibilityDecreasing,Skovorodin2019SuppressionSecondary}

It is partly because of the likely need to include these additional effects in any physical system that our emphasis in this paper has been on physically-motivated estimation methods, rather than calculationally-intensive approximate solutions to the diffusion equation.
Understanding the factors that underpin the confinement time scaling should provide a good foundation on which to build more complex theories involving other physical effects.

\acknowledgments{This work was supported by ARPA-E Grant DE-AR0001554.
This work was also supported by the DOE Fusion Energy Sciences Postdoctoral Research Program, administered by the Oak Ridge Institute for Science and Education (ORISE) and managed by Oak Ridge Associated Universities (ORAU) under DOE contract No. DE-SC0014664.}

\section*{Data Availability Statement}
Data sharing is not applicable to this article as no new data were created or analyzed in this study.



\input{relativisticMirrorResub.bbl}

\clearpage

\end{document}

%% file: relativisticMirrorResub.bbl
%